\documentclass[twocolumn,twoside,slac_two]{revtex4}
\usepackage{graphicx}
\usepackage{fancyhdr}
\begin{document}

\title{Observation of the giant radio galaxy M87 at TeV energies with H.E.S.S.}

%

\author{M. Beilicke, R. Cornils, G. Heinzelmann, M. Raue, J. Ripken}
\affiliation{Institut f\"ur Experimentalphysik, Universit\"at Hamburg, 
Luruper Chaussee 149, D-22761 Hamburg, Germany}
\author{W. Benbow, D. Horns}
\affiliation{Max-Planck-Institut f\"ur Kernphysik, P.O. box 103980, Heidelberg,Germany}
\author{M. Tluczykont}
\affiliation{Laboratoire Leprince-Ringuet, IN2P3/CNRS, Ecole Polytechnique, F-91128 Palaiseau, France}
\author{\vspace{0.2cm}for the H.E.S.S. collaboration}

\begin{abstract}

The giant radio galaxy M\,87 was observed at TeV energies with the
Cherenkov telescopes of the H.E.S.S. collaboration (High Energy
Stereoscopic System). The observations have been performed in the year
2003 during the comissioning phase and in 2004 with the full four
telescope setup. The observations were motivated by the measurement of the
HEGRA collaboration which reported a $4.7 \, \sigma$ excess of TeV
$\gamma$-rays from the direction of M\,87. The results of the H.E.S.S.
observations -- indicating a possible variability of TeV $\gamma$-ray
emission from M\,87 (compared to the HEGRA result) -- are presented.

\end{abstract}

\maketitle

\thispagestyle{fancy}


\section{Introduction}

The giant radio galaxy M\,87 is located at a distance of
$\sim16\,\mathrm{Mpc}$ ($z=0.00436$) in the Virgo cluster of galaxies. The
angle between the parsec scale plasma jet -- well studied at radio,
optical and X-ray wavelengths -- and the observer's line of sight has been
estimated to be in the order of $20^{\circ}-40^{\circ}$. The mass of the
black hole in the center of M\,87 is of the order of $2-3 \, \cdot \,
10^{9}\,M_{\odot}$. M\,87 is discussed to be a powerful accelerator of
high energy particles, possibly even up to the highest energies
\cite{M87_UHECR_1, M87_UHECR_2}. This makes M\,87 an interesting candidate
for TeV $\gamma$-ray emission. M\,87 was observed with the HEGRA
stereoscopic telescope system in 1998/1999 for a total of $77\,\mathrm{h}$
(after quality cuts) above an energy threshold of $730\,\mathrm{GeV}$. An
excess of TeV $\gamma$-rays has been found with a significance of
$4.7\,\sigma$ \cite{HEGRA_M87_1, HEGRA_M87_2}. The integral flux was
calculated to be $3.3\%$ of the flux of the Crab Nebula.

M\,87 is of particular interest for observations at TeV energies: The
large jet angle makes it different from the so far observed TeV emitting
active galactic nuclei (AGN) which are of the blazar type, i.\,e.~with
their plasma jets pointing directly towards the observer. Various models
exist to describe emission of TeV photons from M\,87. Leptonic models
(i.e. inverse Compton scattering) are discussed in \cite{M87_Leptonic},
whereas \cite{M87_Jets} consider the TeV $\gamma$-ray production in large
scale plasma jets. From the experimental view, the TeV $\gamma$-ray
production in large scale jets would be of particular interest since the
extension of the M\,87 jet structure could be resolved at TeV energies
with the typical angular resolution of stereoscopic Cherenkov telescope
arrays of $\le 0.1^{\circ}$ per event. Hadronic models do also exist
\cite{M87_Hadronic_1, M87_Hadronic_2} as well as TeV $\gamma$-ray
production scenarios correlated with the cosmic ray population of the
radio galaxy \cite{M87_Hadronic_3}. Finally, the hypothesis of
annihilating exotic particles (i.e. neutralinos) has been discussed by
\cite{M87_Neutralino}.

Observations with the H.E.S.S. telescopes have been initiated to confirm 
the HEGRA result and to further clarify the origin of the TeV $\gamma$-ray 
emission,

\section{The H.E.S.S. experiment}

\begin{figure*}[t]
\centering
\includegraphics[width=0.98\textwidth]{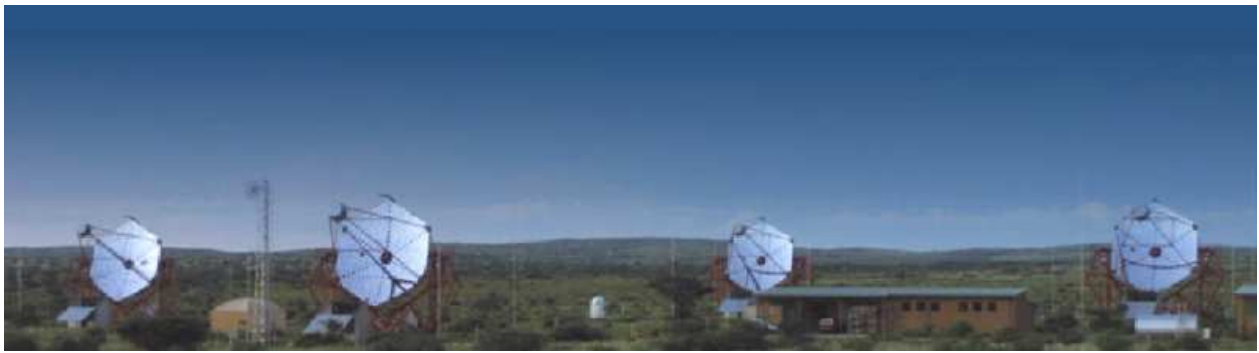}
\caption{The four imaging atmospheric Cherenkov telescopes 
(IACTs) operated by the H.E.S.S. collaboration in Namibia. Since December 
2003 the full four telecope array is 
operational.} 
\label{fig:Hess}
\end{figure*}

The High Energy Stereoscopic System (H.E.S.S.) collaboration operates an
array of four imaging atmospheric Cherenkov telescopes optimized for an
energy range between $100 \, \mathrm{GeV}$ and $10\, \mathrm{TeV}$. The
telescopes are located in the Khomas Highlands in Namibia
($23\mathrm{d}\,16'\,18''\,\mathrm{S}$,
$16\mathrm{d}\,30'\,1''\,\mathrm{E}$)  at a height of $1\,800\,\mbox{m}$
above sea level, see Fig.~\ref{fig:Hess}. Each telescope has a
$107\,\mbox{m}^{2}$ tessellated mirror surface \cite{HESS_Mirror1,
HESS_Mirror2} and is equipped with a 960 photomultiplier tube (PMT) camera
with a field of view of $\sim 5^{\circ}$ \cite{HESS_Camera}. The full four
telescope array is operational since December, 2003. Since July 2003 the
telescopes are operated in a coincident mode \cite{HESS_Trigger} assuring
that at least two telescopes record images for each event which is
important for an improved reconstruction of the shower geometry, and
$\gamma$-hadron separation. More information about H.E.S.S. can be found
in \cite{HESS_Status}.

\section{Observations of M\,87 with H.E.S.S.}

M\,87 has been observed with the H.E.S.S. Cherenkov telescopes between
March and May, 2003 and February to May, 2004. The 2003 data were taken
during the comissioning phase of the experiment with only two telescopes.
The stereo events have been merged offline based on their individual
GPS time stamps. The 2004 data were taken with the full four telescope
array with the hardware coincidence trigger. The sensitivity of the full
setup increased by more than a factor of two compared to the sensitivity
of the instrument during the 2003 observation campaign on M\,87. The
average zenith angle of the observations was $\sim 40^{\circ}$ for both
years. Due to technical reasons one of the four telescopes was excluded 
from the analysis in the February/March 2004 observation period
affecting $\sim 9 \, \mathrm{h}$ of the data by a slightly reduced 
sensitivity.

Standard cuts on the data quality (stable weather and detector status)  
have been applied leaving a dead-time corrected observation time of
$13\,\mathrm{h}$ for the 2003 data and $32\,\mathrm{h}$ for the 2004 data.  
After data calibration \cite{HESS_Calibration} and application of image
cleaning tail-cuts Hillas parameters \cite{HILLAS_Parameters} are
calculated for the individual recorded images. The geometric shower
reconstruction (direction, energy, etc.) follows the standard H.E.S.S.  
analysis technique \cite{HESS_Analysis}.

\section{Results}

\begin{figure}[t]
\centering
\includegraphics[width=0.5\textwidth]{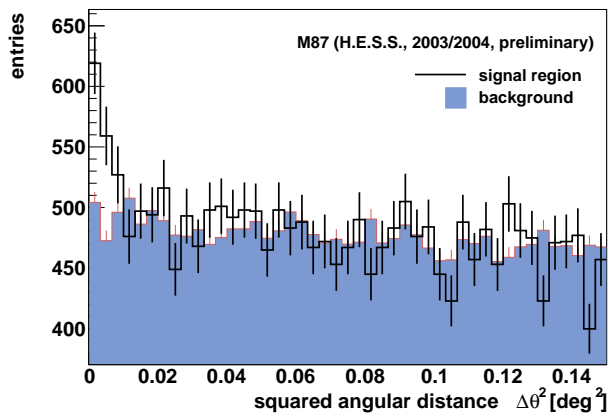}
\caption{Distribution of ON-source events (solid histogram) and
normalized OFF-source events (filled histogram) vs. the squared angular
distance $\Delta\Theta^{2}$ between the reconstructed shower direction and
the nominal object position.} 
\label{fig:OnOff}
\end{figure}

\begin{figure}[t]
\centering
\includegraphics[width=0.5\textwidth]{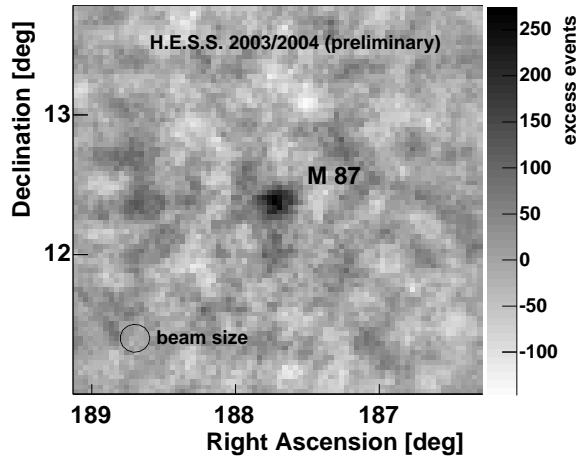}
\caption{The TeV excess sky map showing a $3^{\circ} \times 3^{\circ}$ sky 
region centered around the position of M\,87. The number of events are 
integrated within the optimal point-source angular cut of $\Theta \le 
0.11^{\circ}$ for each of the correlated bins. The background is
estimated using the ring background model. The event-by-event angular 
resolution is indicated by the circle (beamsize).}
\label{fig:SkyMap}
\end{figure}

\begin{figure}[t]
\centering
\includegraphics[width=0.5\textwidth]{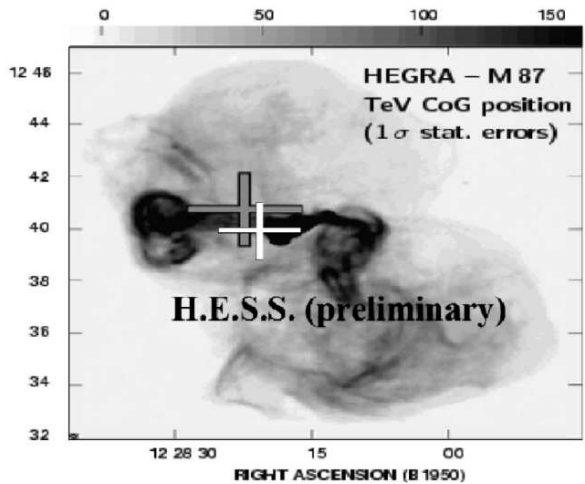}
\caption{The positions of the H.E.S.S. TeV excess (white cross) together 
with the position measured by HEGRA (filled cross) plotted in the radio 
map of M\,87 (adopted from \cite{M87RadioMap}). Within statistical 
errors, both positions are compatible with the M\,87 central region.}
\label{fig:RadioTeVMap}
\end{figure}

Cuts which were optimized on Monte Carlo simulated sources comprising
$10 \%$ of the flux of the Crab Nebula have been applied to the data
including a tight angular cut of $\Delta \Theta^{2} < 0.0125 \,
\mathrm{deg}^{2}$. Although the {\emph software stereo} telescope setup in
2003 would legitimate a separately optimized set of cuts, the same cuts as
for the 2004 data were applied to the 2003 data\footnote{Investigations of
improved analysis techniques optimized for faint sources are underway.}.
The distribution of excess events as a function of the squared angular
distance $\Delta\Theta^{2}$ between the reconstructed shower direction and
the nominal position of M\,87 is shown in Fig.~\ref{fig:OnOff} for the
combined data set. An excess of $216 \pm 49$ events is obtained from the
direction of M\,87 corresponding to a significance of $4.6 \, \sigma$. The
sky map of the combined data set is shown in Fig.~\ref{fig:SkyMap}. The
ring background model was used in which the background is determined from
a ring region with a radius $r = 0.5^{\circ}$ centred around the putative
source position. The position of the TeV excess as measured by H.E.S.S.  
is plotted in the radio map of M\,87 together with the TeV position
reported by HEGRA (see Fig.~\ref{fig:RadioTeVMap}). The TeV excess is
compatible with a point-source and its position was found to be compatible
with the center of the extended structure of M\,87 as well as the position
reported by the HEGRA collaboration within statistical errors. More
observations are needed to reduce the statistical error on the derived
position to further exclude regions within the extended structure of
M\,87.

\begin{figure*}[t]
\centering
\includegraphics[width=0.8\textwidth]{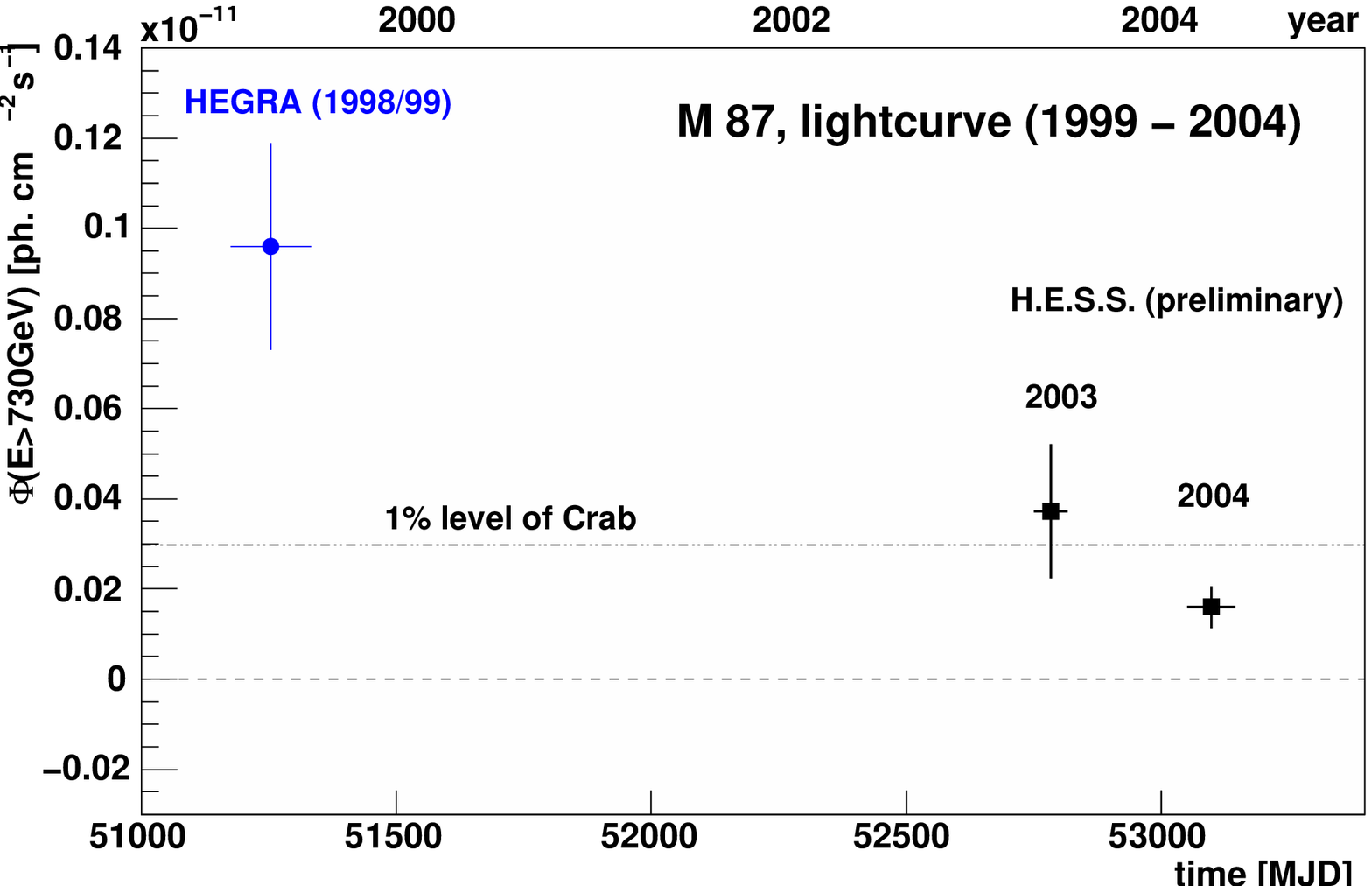}
\caption{The integral photon flux above an threshold of $730\,\mathrm{GeV}$ 
is shown. The recent H.E.S.S. measurement in 2004 is clearly not 
compatible with the HEGRA flux (taken from \cite{HEGRA_M87_1}) which
indicates variability. The $1\,\%$ flux level of the Crab Nebula is also 
indicated underlining the very low (but significant) flux of M\,87 in 2004.}
\label{fig:LightCurve}
\end{figure*}

In order to calculate the integral flux above the energy threshold of the
HEGRA measurement ($730 \, \mathrm{GeV}$) for the 2003 and 2004 H.E.S.S.  
observations a power-law spectrum $\mathrm{d}N/\mathrm{d}E \sim
E^{-\Gamma}$ with a photon index of $\Gamma = 2.9$ (as reported in
\cite{HEGRA_M87_2}) was assumed. The integral flux points are shown in
Fig.~\ref{fig:LightCurve} together with the HEGRA measurement. Note, that
the 2004 flux of M\,87 is below the $1 \%$ flux level of the Crab Nebula.
Although the H.E.S.S. fluxes of 2003 and 2004 seem still to be compatible
-- taken the large statistical error of the 2003 measurement -- the
comparison of the H.E.S.S. 2004 flux with the HEGRA measurement
indicates variable TeV $\gamma$-ray emission from M\,87 on time-scales of 
years.

\section{Summary \& Conclusion}

The giant radio galaxy M\,87 has been observed with H.E.S.S. in 2003
during the comissioning phase and in 2004 with the full four telescope
setup for a total of $45\,\mathrm{h}$ remaining after quality cuts. An
excess of $216 \pm 49$ $\gamma$-ray events has been measured from the
direction of M\,87 with a significance of $4.6\,\sigma$. This confirms the
HEGRA measurement, although on a lower flux level. The TeV excess is
compatible with a point-source and within statistics its position is
located at the center of the extended M\,87 structure. The measured flux
was found to be on the sub $1\,\%$ level of the flux from the Crab Nebula
in the 2004 data which indicates flux variability if compared to the
$3.3\,\%$ flux level in 1998/99 reported by the HEGRA collaboration.

To confirm the indications of variability of the TeV $\gamma$-ray emission
from M\,87 more observations are needed.  Such a result would be
very important since various models for the TeV $\gamma$-ray production in
M\,87 could be ruled out. Mechanisms correlated with cosmic rays
\cite{M87_Hadronic_3}, large scale jet structures \cite{M87_Jets} and
exotic dark matter particle annihilation \cite{M87_Neutralino} could not
explain variability in the TeV $\gamma$-ray emission on these time-scales.
The measurement of an accurate energy spectrum could further help to
reduce the amount of possible models as well as a more precise location of
the emission region; both goals require a measurement with higher event
statistics as currently available by the H.E.S.S. 2003/2004 data set.

M\,87 has been observed in a wide range of the electromagnetic spectrum.  
Simultaneous observations at other wavelengths -- especially in X-rays,
such as the Chandra monitoring of the HST-1 knot in the inner jet region
\cite{Chandra_HST-1} -- are of great importance since a correlation would
further reveal the TeV $\gamma$-ray production mechanism of this AGN which
is the first one not belonging to the blazar class.

\bigskip
\begin{acknowledgments}

The support of the Namibian authorities and of the University of Namibia
in facilitating the construction and operation of H.E.S.S. is gratefully
acknowledged, as is the support by the German Ministry for Education and
Research (BMBF), the Max Planck Society, the French Ministry for Research,
the CNRS-IN2P3 and the Astroparticle Interdisciplinary Programme of the
CNRS, the U.K. Particle Physics and Astronomy Research Council (PPARC),
the IPNP of the Charles University, the South African Department of
Science and Technology and National Research Foundation, and by the
University of Namibia. We appreciate the excellent work of the technical
support staff in Berlin, Durham, Hamburg, Heidelberg, Palaiseau, Paris,
Saclay, and in Namibia in the construction and operation of the equipment.

\end{acknowledgments}

\bigskip 

\begin{thebibliography}{99} 

\bibitem{HESS_Analysis}
Aharonian, F., et al. (H.E.S.S. collab.), 2005,
A\&A, 430, 865

\bibitem{HESS_Mirror1}
Bernl\"ohr, K. et al. (H.E.S.S. collab.), 2003,
Astroparticle Physics, 20, 111

\bibitem{HESS_Mirror2}
Cornils, R. et al. (H.E.S.S. collab.), 2003,
Astroparticle Physics, 20, 129

\bibitem{HESS_Camera}
Vincent, P., Denance, J.-P., Huppert, J.-F., et al. (H.E.S.S. collab.), 2003,
Proc. of the 28$^{th}$ ICRC (Tsukuba), p.2887

\bibitem{HESS_Trigger}
Funk, S., Hermann, G., Hinton, J., et al. (H.E.S.S. collab.), 2004,
Astroparticle Physics, 22/3-4, 285-296

\bibitem{HESS_Status}
{\emph http://www.mpi-hd.mpg.de/hfm/HESS/HESS.html}

\bibitem{HESS_Calibration}
Aharonian, F. et al. (H.E.S.S. collab.), 2004,
Astroparticle Physics, 22, 109

\bibitem{HILLAS_Parameters}
Hillas, A.M.: 1985,
Proc. of 19th ICRC (La Jolla), Vol.3, 445

\bibitem{HEGRA_M87_1}
Aharonian, F., et al. (HEGRA collab.), 2003,
A\&A, 403, L1

\bibitem{HEGRA_M87_2}
G\"otting, N. et al. (HEGRA collab.), 2003,
The European Physical Journal C - Particles and Fields, see astro-ph/0310308

\bibitem{M87_Leptonic}
Bai, J.M., \& Lee, M.G., 2001,
ApJ, 549, L173

\bibitem{M87_Jets}
Stawarz, L. et al., 2003,
ApJ, 597, 186-201

\bibitem{M87_Hadronic_1}
Protheroe, R.J. et al., 2003,
Astroparticle Physics, Vol.19, Issue 4, 559

\bibitem{M87_Hadronic_2}
Reimer, A., et al., 2004,
A\&A 419, 89-98

\bibitem{M87_Hadronic_3}
Pfrommer, C. \& Enslin, T.A., 2003,
A\&A, 407, L73

\bibitem{M87_Neutralino}
Baltz et al., 1999,
Physical Review D, 61, 023514

\bibitem{M87_UHECR_1}
Ginzburg, V. L. \& Syrovatskii, S. L., 1964,
"The origin of cosmic rays", Pergamon Press, Oxford

\bibitem{M87_UHECR_2}
Biermann, et al., 2000,
Nucl. Phys. B, Proc. Suppl., 87, 417

\bibitem{M87RadioMap}
Owen, F.N., Ledlow, M.J., Eilek, J.A., et al., 2000,
Proc. of The Universe at Low Radio Frequencies, ASP Conf. Ser., 199, 
{\emph see astro-ph/0006152}

\bibitem{Chandra_HST-1}
Harris, D.E. et al., 2003,
ApJ, 586, L41

\end{thebibliography}

\end{document}